\documentclass[Physsubmission,Phys]{SciPost}
\usepackage{graphicx}
\usepackage{media9}

\usepackage{tikz,varwidth}
\usetikzlibrary{shapes.geometric}
\usetikzlibrary{arrows.meta,shapes,automata,petri,positioning,calc,decorations.markings}

\hypersetup{
    colorlinks,
    linkcolor={red!50!black},
    citecolor={blue!50!black},
    urlcolor={blue!80!black}
}

\usepackage[bitstream-charter]{mathdesign}
\DeclareSymbolFont{usualmathcal}{OMS}{cmsy}{m}{n}
\DeclareSymbolFontAlphabet{\mathcal}{usualmathcal}

\begin{document}

\begin{center}{\Large \textbf{
Centre Vortex Structure of QCD-Vacuum Fields and Confinement
}}\end{center}

\begin{center}
Derek Leinweber\textsuperscript{1$\star$},
James Biddle\textsuperscript{1} and
Waseem Kamleh\textsuperscript{1}
\end{center}

\begin{center}
{\bf 1} Centre for the Subatomic Structure of Matter, Department of Physics,\\
        University of Adelaide, SA 5005, Australia
\\
* derek.leinweber@adelaide.edu.au
\end{center}

\begin{center}
\today
\end{center}


\definecolor{palegray}{gray}{0.95}
\begin{center}
\colorbox{palegray}{
  \begin{minipage}{0.95\textwidth}
    \begin{center}
    {\it  XXXIII International (ONLINE) Workshop on High Energy Physics \\
          “Hard Problems of Hadron Physics:  Non-Perturbative QCD \& Related Quests”}\\
    {\it November 8-12, 2021} \\
    \doi{10.21468/SciPostPhysProc.?}\\
    \end{center}
  \end{minipage}
}
\end{center}

\section*{Abstract}
{\bf
The non-trivial ground-state vacuum fields of QCD form the foundation of matter.  Using modern
visualisation techniques, this presentation examines the microscopic structure of these fields.  Of
particular interest are the centre vortices identified within the ground-state fields of lattice
QCD.  Our current focus is on understanding the manner in which dynamical fermions in the QCD
vacuum alter the centre-vortex structure.  The impact of dynamical fermions is significant and
provides new insights into the role of centre vortices in underpinning both confinement and
dynamical chiral symmetry breaking in QCD.
}

\vspace{10pt}
\noindent\rule{\textwidth}{1pt}
\tableofcontents\thispagestyle{fancy}
\noindent\rule{\textwidth}{1pt}
\vspace{10pt}

\section{Introduction}
\label{sec:intro}

Lattice QCD calculations have been instrumental in revealing the fundamental role of centre
vortices~\cite{tHooft:1977nqb,tHooft:1979rtg,DelDebbio:1996lih,Faber:1997rp,%
  DelDebbio:1998luz,Bertle:1999tw,Faber:1999gu,Engelhardt:1999xw,%
  Bertle:2000qv,Greensite:2003bk,Engelhardt:2003wm,Greensite:2016pfc} in the ground-state vacuum
fields in governing the confinement of quarks.

By identifying centre vortices and then removing them from QCD ground-state fields, a deep
understanding of their contributions has been developed.  Removal of centre vortices from the
ground-state fields results in a loss of dynamical mass generation and restoration of chiral
symmetry~\cite{Trewartha:2015nna,Trewartha:2017ive}, a loss of the string
tension~\cite{Langfeld:2003ev,Bowman:2010zr} and a suppression of the infrared enhancement of the
Landau-gauge gluon propagator~\cite{Langfeld:2001cz,Bowman:2010zr,Biddle:2018dtc}.

One can also examine the role of the centre vortices alone.  Remarkably centre vortices produce
both a linear static quark potential~\cite{Langfeld:2003ev,OCais:2008kqh,Trewartha:2015ida} and
infrared enhancement in the Landau-gauge gluon propagator.  The planar vortex density of
centre-vortex degrees of freedom scales with the lattice spacing providing an well defined
continuum limit~\cite{Langfeld:2003ev}.  These results elucidate strong connections between centre
vortices and confinement.

A connection between centre vortices and instantons was identified through gauge-field
smoothing~\cite{Trewartha:2015ida}.  An understanding of the phenomena linking these degrees
of freedom was illustrated in Ref.~\cite{Biddle:2019gke}.  In addition, centre vortices have been
shown to give rise to mass splitting in the low-lying hadron spectrum
~\cite{Trewartha:2017ive,Trewartha:2015nna,OMalley:2011aa}.

Still, the picture is not perfect. The vortex-only string tension obtained from pure Yang-Mills
lattice studies has been consistently shown to be about $\sim 62\%$ of the full string
tension. Moreover, upon removal of centre vortices the gluon propagator shows a remnant of infrared
enhancement~\cite{Biddle:2018dtc}.  In the pure gauge sector, the removal of long-distance
non-perturbative effects via centre-vortex removal is not perfect.

Here we turn our attention to understanding the impact of dynamical fermions on the centre-vortex
structure of QCD ground-state fields.  We will illustrate the differences in the microscopic
structure and reveal how the change in structure affects the static quark potential and the
Landau-gauge gluon propagator.  We find the introduction of dynamical fermions brings the
phenomenology of centre vortices much closer to a perfect encapsulation of the salient features of
QCD.

\section{Centre Vortex Identification}

In identifying centre vortices one commences with a gauge fixing procedure which brings the lattice
link variables as close as possible to the identity times a phase.  Here, the original Monte-Carlo
generated configurations are considered.  They are gauge transformed directly to Maximal Centre
Gauge \cite{DelDebbio:1996mh,Langfeld:1997jx,Langfeld:2003ev}, without
preconditioning~\cite{Cais:2008za}.  The brings the lattice link variables $U_\mu(x)$ close to the
centre elements of $SU(3)$,
\begin{equation}
Z = \exp \left ( 2 \pi i\, \frac{m}{3} \right ) \, \mathbf{I}, \textrm{ with } m = -1, 0, 1.
\label{CentreSU3}
\end{equation}
One considers gauge transformations $\Omega$ such that,
\begin{equation}
\sum_{x,\mu} \,  \left | \mathrm{tr}\, U_\mu^\Omega(x) \, \right |^2 \stackrel{\Omega}{\to}
\mathrm{max} \, ,
\label{GaugeTrans}
\end{equation}
and then projects the link variables to the centre
\begin{equation}
U_\mu(x) \to Z_\mu(x) \textrm{ where }
Z_\mu(x) = \exp \left ( 2 \pi i\, \frac{m_\mu(x)}{3} \right
)\mathbf{I} \, ,
\label{UtoZ}
\end{equation}
and $m_\mu(x) = -1, 0, 1.$

The product of these centre-projected links around an elementary $1\times 1$ square on the lattice
reveals the centre charge associated with that plaquette.  The centre-line of an extended vortex in three
dimensions is identified by tracing the presence of nontrivial centre charge, $z$, through the
spatial lattice
\begin{equation}
z = \prod_\Box Z_\mu(x) = \exp \left ( 2 \pi i\, \frac{m}{3} \right ) \, .
\label{CentreCharge}
\end{equation}
A right-handed ordering of the dimensions is selected in calculating and illustrating the centre
charge.  If $z=1$, no vortex pierces the plaquette. If $z \neq 1$ a vortex with charge $z$ pierces
the plaquette.  We refer to the centre charge of a vortex via the value of $m = \pm 1$.

\section{Centre Vortex Visualisation}

\begin{figure}[t]
\begin{center}
\resizebox{!}{3cm}{%
\begin{tikzpicture}[scale=0.9]
\begin{scope}[very thick,decoration={
    markings,
    mark=at position 0.5 with {\arrow[scale=2]{stealth}}}
    ]
  \draw[line width=1.0,postaction={decorate}](1.5,-1.5)-- node[left]{$Z_y(n+\hat x)\ $} (3.25,1.5)node(g){};
  \draw[line width=1.0,postaction={decorate}](3.25,1.5)-- node[above]{${}\ \ Z_x^*(n+\hat y)$} (-1.5,1.5);
  \draw[line width=1.0,postaction={decorate}](-1.5,1.5)-- node[left]{$Z_y^*(n)$}(-4,-1.5);

  \draw (-0.3,-2) node(a){}
  -- (0.1,-2) node(b){}   
  -- (-0.1,2) node(c){}   
  -- cycle;               
  \fill[blue] (a.center) -- (b.center) -- (c.center);

  \draw[line width=1.0,postaction={decorate}](-4,-1.5)-- node[above]{$Z_x(n)$}(1.5,-1.5)node(f){};

  \draw[line width=1.0,-{Latex[length=2mm]}](3.5,0)--(4.5,0.0)node[right]{\large $x$};
  \draw[line width=1.0,-{Latex[length=2mm]}](3.5,0)--(3.9,0.8)node[right]{\large $y$};
  \draw[line width=1.0,-{Latex[length=2mm]}](3.5,0)--(3.5,1.0)node[above]{\large $z$};
  \end{scope}
\end{tikzpicture}
}
\resizebox{!}{3cm}{%
\begin{tikzpicture}[scale=0.9]
\begin{scope}[very thick,decoration={
    markings,
    mark=at position 0.5 with {\arrow[scale=2]{stealth}}}
    ]
  \draw[line width=1.0,postaction={decorate}](1.5,-1.5)-- node[left]{$Z_y(n+\hat x)\ $} (3.25,1.5)node(g){};
  \draw[line width=1.0,postaction={decorate}](3.25,1.5)-- node[above]{\quad $Z_x^*(n+\hat y)$} (-1.5,1.5);
  \draw[line width=1.0,postaction={decorate}](-1.5,1.5)-- node[left]{$Z_y^*(n)$}(-4,-1.5);

  \draw (-0.3,2) node(a){}
  -- (0.1,2) node(b){}
  -- (-0.1,-2)node(c){}
  -- cycle;
  \fill[blue] (a.center) -- (b.center) -- (c.center);

  \draw[line width=1.0,postaction={decorate}](-4,-1.5)-- node[above]{$Z_x(n)$}(1.5,-1.5)node(f){};

  \draw[line width=1.0,-{Latex[length=2mm]}](3.5,0)--(4.5,0.0)node[right]{\large $x$};
  \draw[line width=1.0,-{Latex[length=2mm]}](3.5,0)--(3.9,0.8)node[right]{\large $y$};
  \draw[line width=1.0,-{Latex[length=2mm]}](3.5,0)--(3.5,1.0)node[above]{\large $z$};
  \end{scope}
\end{tikzpicture}
}
\end{center}
\vspace{-18pt}
\caption{{\bf Illustrating nontrivial centre charge via a jet.} (left) An $m = +1$ vortex with
  centre charge $z = \exp(2\pi i / 3)$ is rendered as a jet pointing in the $+\hat z$ direction.
  (right) An $m = -1$ vortex with centre charge $z = \exp(-2\pi i / 3)$ is rendered as a jet in the
  $-\hat z$ direction.  }
\label{fig:jets}
\end{figure}
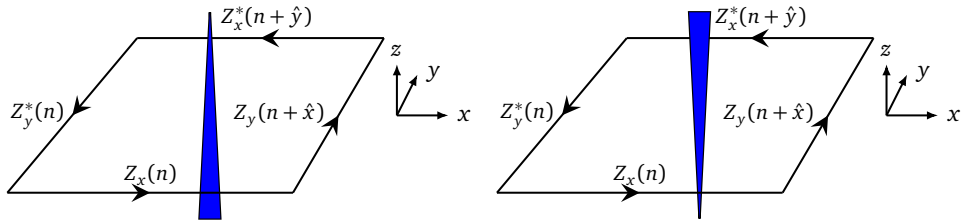

The centre lines of extended vortices are illustrated on the dual lattice by rendering a jet
piercing the plaquette producing the nontrivial centre charge.  The orientation of the jet follows
the right-handed coordinate system.  Figure \ref{fig:jets} provides an illustration of this
assignment. For example, with reference to Eq.~(\ref{CentreCharge}), an $m = +1$ vortex in the
$x$-$y$ plane is plotted in the $+\hat z$ direction as a blue jet.  Similarly, an $m = -1$ vortex
in the $x$-$y$ plane is plotted in the $-\hat z$ direction.
As the centre charge transforms to its complex conjugate under permutation of the two dimensions
describing the plaquette, the centre charge can be thought of as the directed flow of charge $z =
\exp(2\pi i / 3)$.

Our current focus is to understand the impact of dynamical-fermion degrees of freedom on the
centre-vortex structure of a gluon field.  Here we consider the PACS-CS $(2+1)$-flavour full-QCD
ensembles~\cite{Aoki:2008sm}, made available through the ILDG~\cite{Beckett:2009cb}. These $32^3
\times 64$ lattice ensembles employ a renormalisation-group improved Iwasaki gauge action with
$\beta = 1.90$ and non-perturbatively ${\cal O}(a)$-improved Wilson quarks, with $C_{\rm SW} =
1.715$.  In this section, their lightest $u$- and $d$-quark-mass ensemble identified by a pion mass
of 156 MeV~\cite{Aoki:2008sm} is considered.  The scale is set using the Sommer parameter with $r_0
= 0.4921$ fm providing a lattice spacing of $a=0.0933$ fm~\cite{Aoki:2008sm}.

For comparison, a matched $32^3 \times 64$ pure-gauge ensemble has been generated using the same
improved Iwasaki gauge action with $\beta = 2.58$ providing a Sommer-scale spacing of $a = 0.100$
fm.  This spacing facilitates comparisons with all the PACS-CS ensembles.

\begin{figure}[p]
\null\hspace{-0.3cm}
\includemedia[
        noplaybutton,
	3Dtoolbar,
	3Dmenu,
	3Dviews=U3D/Primary-Secondary-PG-28.vws,
	3Dcoo  = 16 16 32, 
        3Dc2c=0.245114266872406 0.8673180341720581 0.43321868777275085,
	3Droo  = 110.0,    
	3Droll =-98.1,     
	3Dlights=Default,  
	width=1.04\textwidth,  
]{\includegraphics{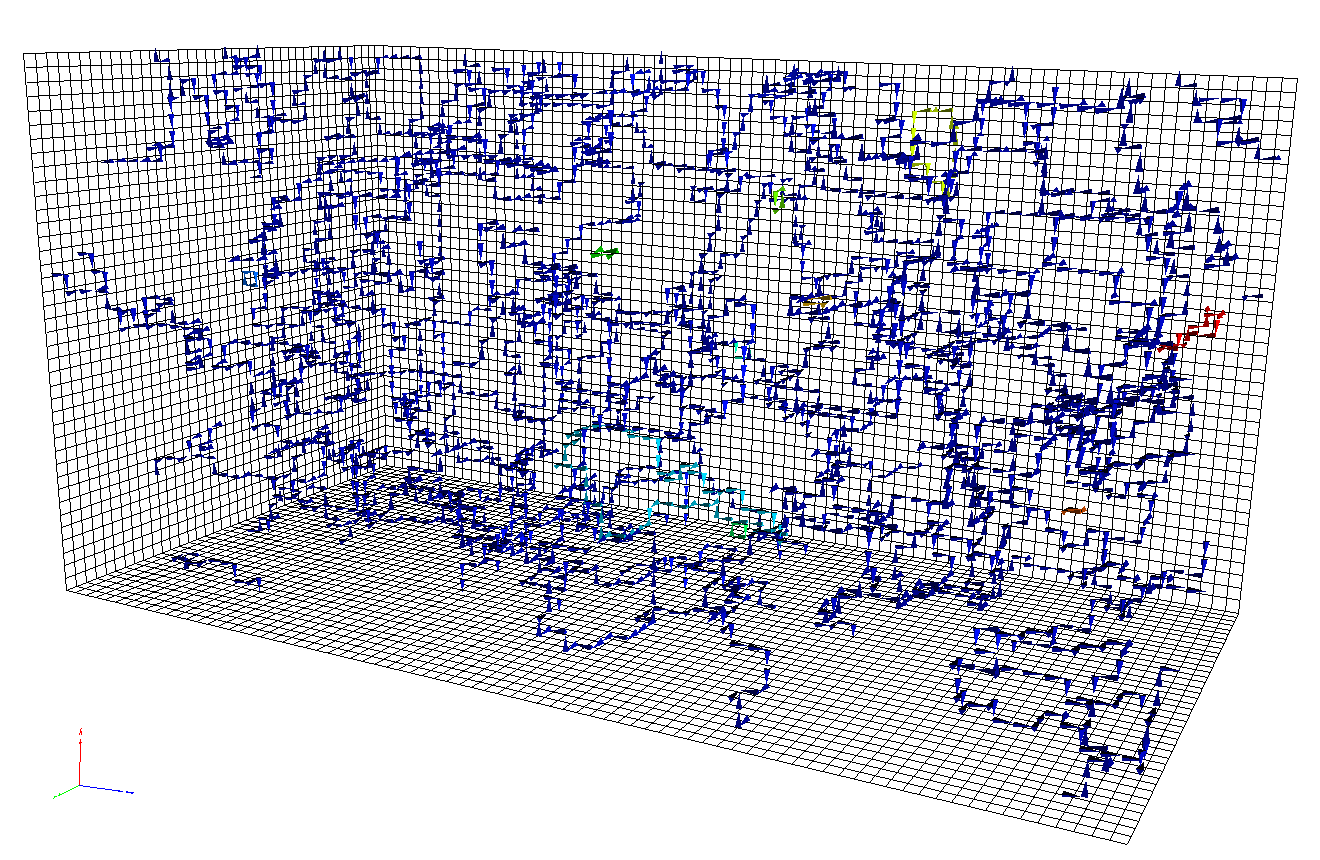}}{U3D/Primary-Secondary-PG-28.u3d}
\vspace{-24pt}
\caption{
  {\bf The centre-vortex structure of a ground-state vacuum field
    configuration in pure SU(3) gauge theory.}
  ({\sl Click to activate.})
  The flow of $+1$ centre charge through a gauge field is illustrated by the jets.
  Blue jets are used to illustrate the single percolating vortex structure, while
  other colours illustrate smaller structures.
\label{Primary-Secondary-PG-28.u3d}
}
\begin{center}
\includemedia[
        noplaybutton,
	3Dtoolbar,
	3Dmenu,
	3Dviews=U3D/Primary-Secondary-DF-18.vws,
	3Dcoo  = 16 16 32, 
        3Dc2c=0.245114266872406 0.8673180341720581 0.43321868777275085,
	3Droo  = 110.0,    
	3Droll =-98.1,     
	3Dlights=Default,  
	width=\textwidth,  
]{\includegraphics{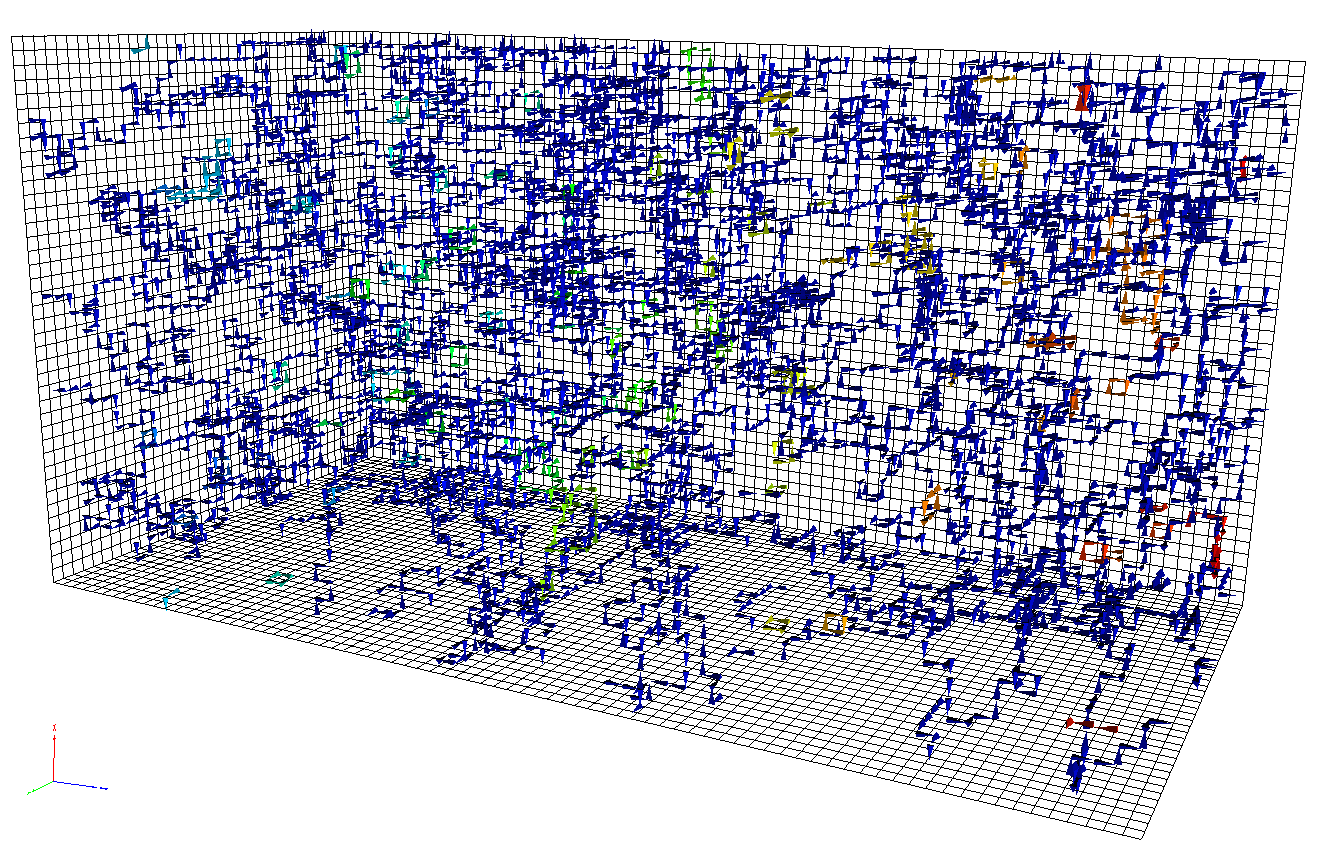}}{U3D/Primary-Secondary-DF-18.u3d}
\end{center}
\vspace{-18pt}
\caption{
  {\bf The centre-vortex structure of a ground-state vacuum field
    configuration in dynamical 2+1 flavour QCD.}
  ({\sl Click to activate.})
  The flow of $+1$ centre charge through a gauge field is illustrated by the jets.
  Symbols are as described in Fig.~\ref{Primary-Secondary-PG-28.u3d}.
\label{Primary-Secondary-DF-18.u3d}
}
\end{figure}

The centre-vortex structure of pure-gauge and dynamical-fermion ground-state vacuum fields is
illustrated in Figs.~\ref{Primary-Secondary-PG-28.u3d} and \ref{Primary-Secondary-DF-18.u3d}
respectively.  The vortex flow displays a rich structure.  One observes the continuous flow of
centre charge and the presence of monopole or anti-monopole contributions, where three jets
emerge from or converge to a point.  We refer to the latter as branching points in general.  Upon
introducing dynamical fermions, the structure becomes more complicated, both in the abundance of
nontrivial centre charge and in the increased abundance of branching points.

These figures provide interactive illustrations which
can be activated in Adobe Reader\footnote{Open this pdf document in
  Adobe Reader 9 or later.  Linux users can install
  \href{ftp://ftp.adobe.com/pub/adobe/reader/unix/9.x/9.4.1/enu/}{Adobe
    acroread version 9.4.1}, the last edition to have full 3D support.
  From the ``Edit'' menu, select ``Preferences...'' and ensure ``3D \&
  Multimedia'' is enabled and ``Enable double-sided rendering'' is
  selected.}  by clicking on the image.  Once activated, click and
drag to rotate, Ctrl-click to translate, Shift-click or mouse wheel to
zoom, and right click to access the ``Views'' menu.  Several views
have been created to facilitate and inspection of the centre-vortex
structure.

Both Figs.~\ref{Primary-Secondary-PG-28.u3d} and \ref{Primary-Secondary-DF-18.u3d} contain a
percolating vortex cluster, a characteristic feature of the confining
phase~\cite{Engelhardt:1999fd}.  These illustrations are representative of the ensemble in that the
vortex vacuum is typically dominated by a single large percolating cluster.  This single large
cluster is accompanied by several smaller loops or loop clusters.  However, the most important
observation is how dynamical fermions significantly increase the number of vortices observed.

For an ensemble of 200 configurations with $32$
three-dimensional volume slices each, the average number of vortices
composing the primary cluster in these $32^2 \times 64$ spatial slices
is
$3,277\, \pm 156$ vortices in the pure gauge theory, versus
$5,924\, \pm 239$ vortices in full QCD.
Since there are $32^2 \times 64 \times 3 = 196,608$ spatial
plaquettes on these lattices, the presence of a vortex
is a relatively rare occurrence.

\begin{figure}[p]
\begin{center}
\includemedia[
        noplaybutton,
	3Dtoolbar,
	3Dmenu,
	3Dviews=U3D/Secondary-PG-28.vws,
	3Dcoo  = 16 16 32, 
        3Dc2c=0.245114266872406 0.8673180341720581 0.43321868777275085,
	3Droo  = 110.0,    
	3Droll =-98.1,     
	3Dlights=Default,  
	width=1.04\textwidth,  
]{\includegraphics{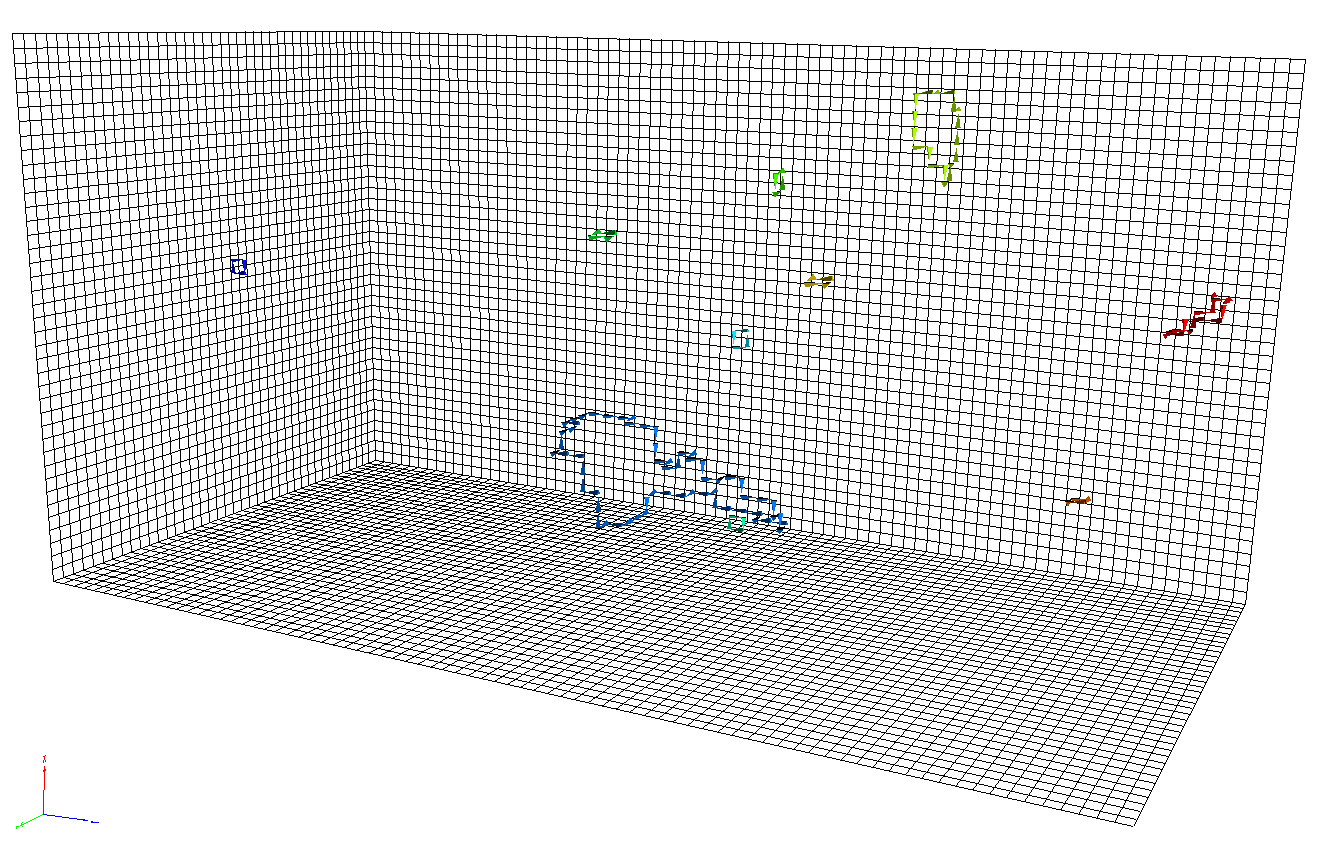}}{U3D/Secondary-PG-28.u3d}
\end{center}
\vspace{-24pt}
\caption{
  {\bf The centre-vortex structure of secondary loops in a ground-state vacuum field
    configuration of pure SU(3) gauge theory.}
  ({\sl Click to activate.})
  The flow of $+1$ centre charge in the secondary loops -- left behind as the
  single percolating structure is removed -- is illustrated.
  %
\label{Secondary-PG-28.u3d}
}
\null\hspace{-0.25cm}
\includemedia[
        noplaybutton,
	3Dtoolbar,
	3Dmenu,
	3Dviews=U3D/Secondary-DF-18.vws,
	3Dcoo  = 16 16 32, 
        3Dc2c=0.245114266872406 0.8673180341720581 0.43321868777275085,
	3Droo  = 110.0,    
	3Droll =-98.1,     
	3Dlights=Default,  
	width=1.05\textwidth,  
]{\includegraphics{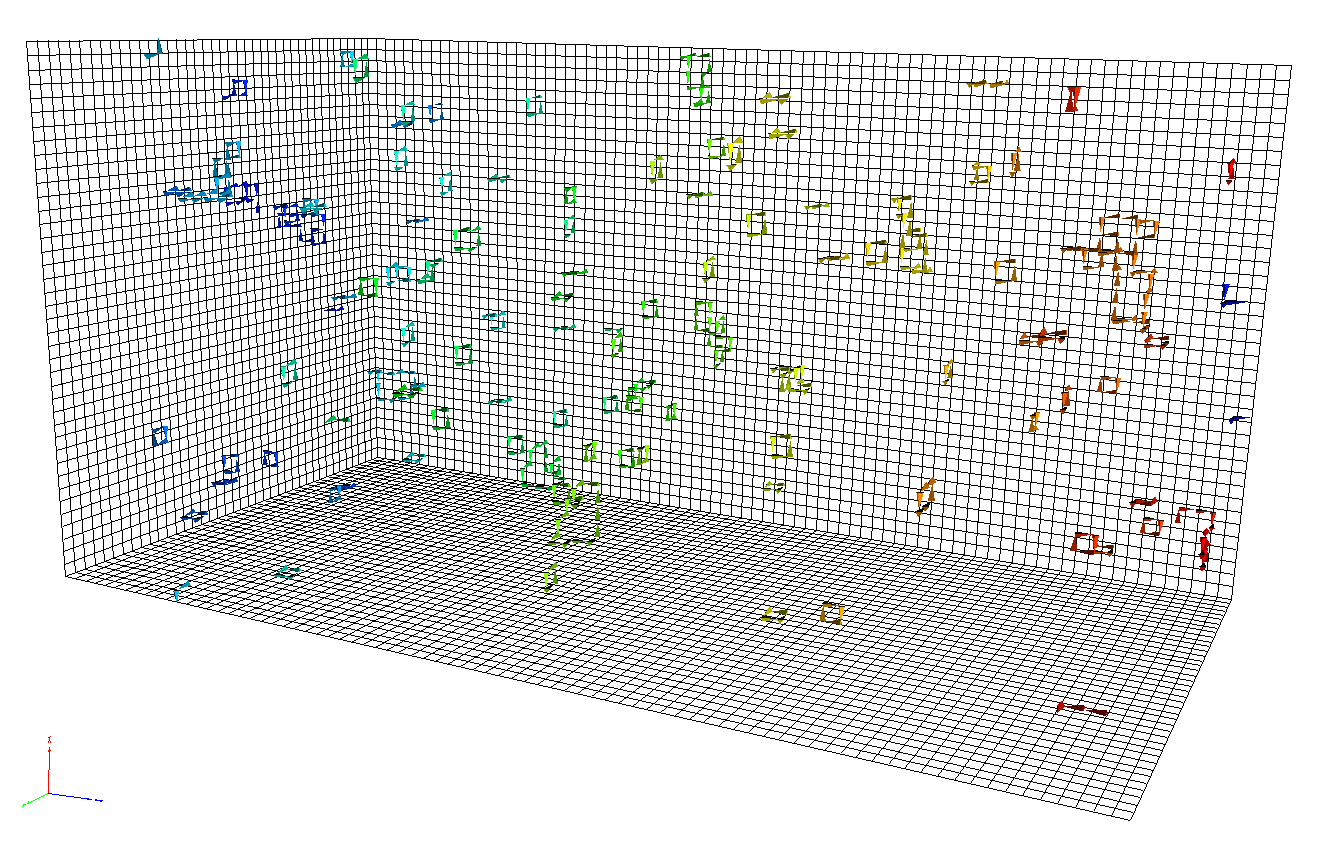}}{U3D/Secondary-DF-18.u3d}
\vspace{-24pt}
\caption{
  {\bf The centre-vortex structure of secondary loops in a ground-state vacuum field
    configuration of dynamical 2+1 flavour QCD.}
  ({\sl Click to activate.})
  The flow of $+1$ centre charge in the secondary loops is
  illustrated.
  %
\label{Secondary-DF-18.u3d}
}
\end{figure}

Similarly, Figs. \ref{Secondary-PG-28.u3d} and \ref{Secondary-DF-18.u3d} illustrate the secondary
loop structures left behind as one removes the single large percolating structure.  Again, the
introduction of dynamical fermions increases the complexity of the structure through a
proliferation of branching points (or monopoles~\cite{Spengler:2018dxt}).  Figure
\ref{Secondary-DF-18.u3d} contains many views in the drop-down menu to facilitate the observation
of this complexity.

\section{Static Quark Potential}

With an understanding the impact of dynamical-fermion degrees of freedom on the centre-vortex
structure of ground-state vacuum fields, we turn our attention to confinement as realised in the
static quark potential.  The results presented here are supported by complimentary studies of the
nonperturbative gluon propagator.

The static quark potential is accessed via consideration of the expectation value of Wilson loops,
$\left\langle W(r, t) \right\rangle$, with spatial separation $r$ and temporal extent $t$,
\begin{equation}
  \label{eq:1}
\langle W(r, \, t) \rangle = \sum_\alpha \lambda^{\alpha}\!(r)\, \exp \left(-V^{\alpha}\!(r)\, t\right)\, .
\end{equation}
The relevant static quark potential is given by the lowest $\alpha = 0$ state.  We use a
variational analysis of several spatially-smeared sources to isolate this state.

With knowledge of the vortex content of a configuration, contained in $Z_\mu(x)$ of
Eq.~(\ref{UtoZ}), we can analyse two vortex-modified ensembles in addition to the original
untouched configuration, $U_\mu(x)$.  We refer to these as the vortex-only, $Z_\mu(x)$, and
vortex-removed, $Z_\mu^\dagger(x)\, U_\mu(x)$, ensembles.

The static quark potential for the original untouched configurations is expected to follow a
Cornell potential
\begin{equation}
  \label{eq:2}
  V(r) = V_0 - \frac{\alpha}{r} + \sigma\, r \, ,
\end{equation}
composed of a Coulomb term, dominant at short distances, and a linear term, dominant at large
distances.  As centre vortices are anticipated to encapsulate the non-perturbative long-range
physics, the vortex-only results should give rise to a linear
potential~\cite{Greensite:2016pfc,DelDebbio:1998luz,Dosch:1988ha}.  On the other hand, the
vortex-removed results are expected to capture the short-range behaviour.
To analyse the linearity of the potential at large distances we plot a sliding local linear fit to
the potential with extent $r \pm \frac{3}{2}a$.

\begin{figure}[p]
  \centering
  \includegraphics[width=0.8\textwidth]{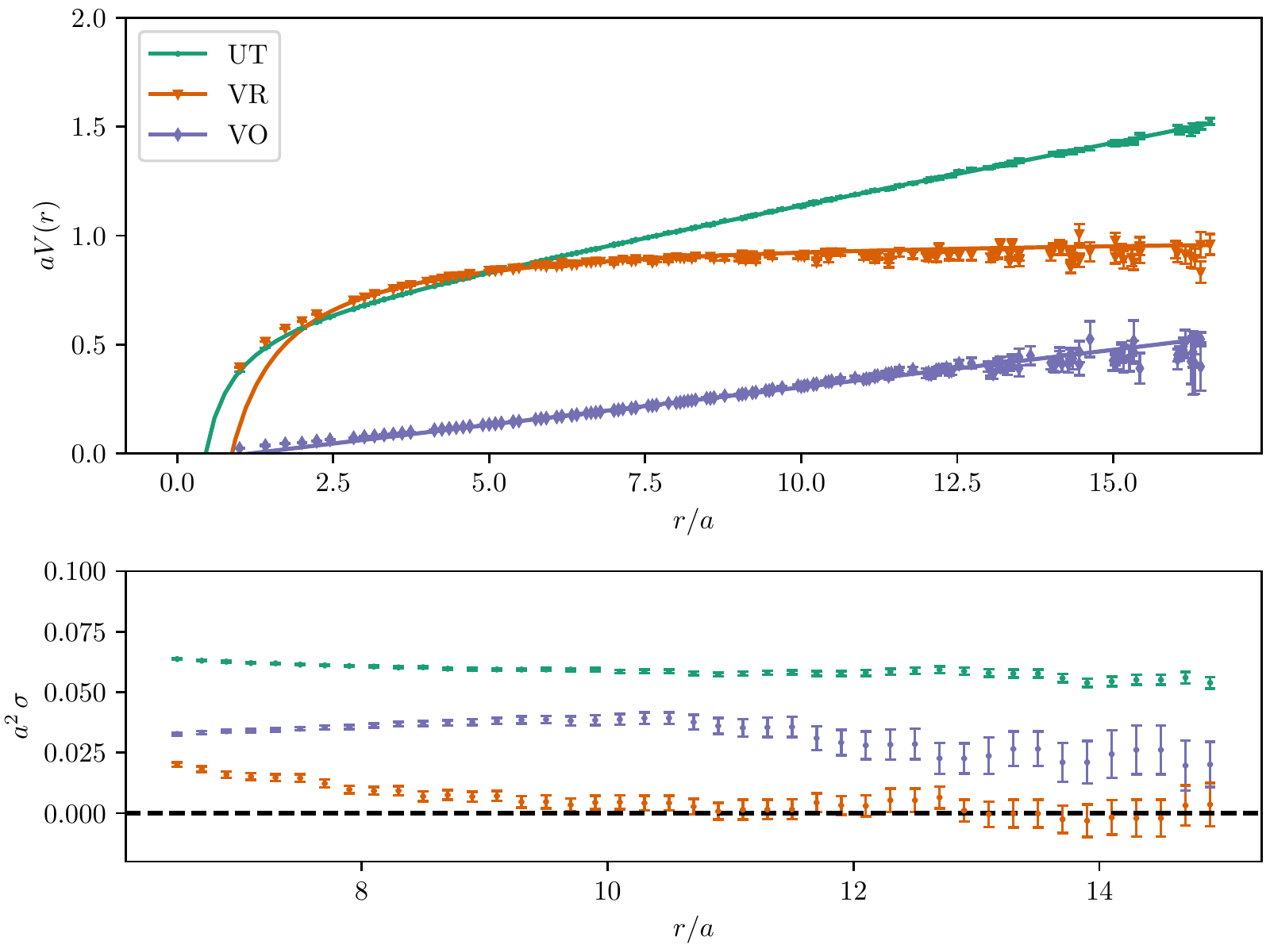}
  \caption{\label{fig:sqp-PG} {\bf The static quark potential as calculated on the original untouched
    and vortex-modified pure-gauge ensembles.} The lower plot shows the local slope of the potentials
    at position $r$ obtained from a linear fit with extent $r \pm \frac{3}{2}a$.  }
\end{figure}

\begin{figure}[p]
  \centering
  \includegraphics[width=0.8\textwidth]{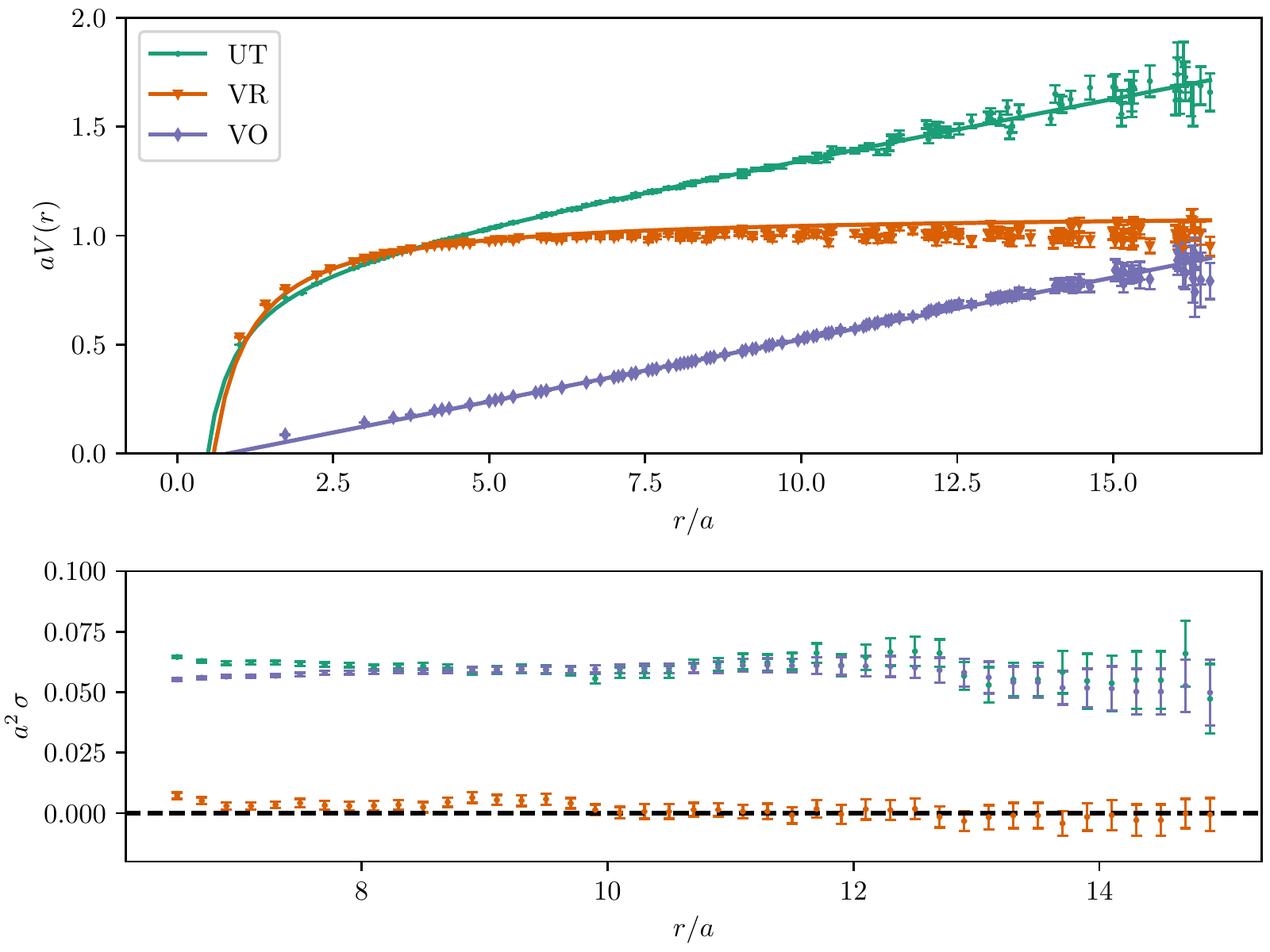}
  \caption{\label{fig:sqp-heavy} {\bf The static quark potential as calculated on the
    vortex-modified dynamical-fermion ensembles, corresponding to a pion mass of $701$ {MeV}.}
    Details are as in Fig.~\ref{fig:sqp-PG}.}
\end{figure}

We commence with preliminary results for the pure-gauge ensemble illustrated in
Fig.~\ref{fig:sqp-PG}.  Qualitatively, centre vortices account for the long-distance physics. The
lower plot illustrates how removal of centre vortices completely removes the long-range
potential. However, the phenomenology is not perfect.  The value of the string tension produced in
the vortex-only analysis is once again only $~60$ \% of the original string tension.

Upon introducing dynamical fermions with light quark masses corresponding to a pion mass of $m_{\pi}
= 701$ {MeV}, the preliminary results shown in Fig.~\ref{fig:sqp-heavy} are observed.  Comparing
with the pure-gauge sector of Fig.~\ref{fig:sqp-PG}, we observe a screening of the original string
tension with the introduction of dynamical fermions, in accord with expectations.  Again, the effect
of vortex removal is to remove confinement. The sliding average lies in excellent agreement with 0
at large distances.

While the centre-vortex phenomenology is similar to the pure gauge sector, this time the
vortex-only string tension is in excellent agreement with the original untouched ensemble.  This is
illustrated in Fig.~\ref{fig:sqp-heavy}, in the lower plot where the local slopes of the untouched
and vortex-only ensembles agree at large distances.  This new agreement arises from significant
modifications in the centre-vortex structure of ground state fields induced by dynamical fermions,
even at relatively large quark masses.

\section{Conclusion}

In summary, centre-vortex structure is complex.  Each ground-state configuration is dominated by a
long-distance percolating centre-vortex structure.  In $SU(3)$ gauge field theory, a proliferation
of branching points is observed, with further enhancement as light dynamical fermion degrees of
freedom are introduced in simulating QCD.  There is an approximate doubling in the number of
nontrivial centre charges in the percolating vortex structure as one goes from the pure-gauge
theory to full QCD.  An enhancement in the number of small vortex paths is also observed upon
introducing dynamical fermions.  Increased complexity in the vortex paths is also observed as the
number of monopole-antimonopole pairs is significantly increased with the introduction of dynamical
fermions.  In short, dynamical-fermion degrees of freedom radically alter the centre-vortex
structure of the ground-state vacuum fields.

With regard to the static quark potential and confinement, we find that centre vortices now
quantitatively capture the string tension in full QCD, unlike the pure-gauge sector.  This
represents a significant advance in centre-vortex phenomenology.  Moreover, vortex removal
continues to eliminate the long distance potential.  These encouraging results are also reflected
in more recent studies of the gluon propagator in full QCD.  In summary, the results presented here
show a significant advance in the ability of centre vortices to capture the salient nonperturbative
features of QCD.

\section*{Acknowledgements}

We thank the PACS-CS Collaboration for making their 2 +1
flavour configurations available via the International Lattice Data Grid
(ILDG).

\paragraph{Funding information}
This research was undertaken with the assistance of resources from the National Computational
Infrastructure (NCI), provided through the National Computational Merit Allocation Scheme and
supported by the Australian Government through Grant No. LE190100021 via the University of Adelaide
Partner Share. This research is supported by Australian Research Council through Grants
No. DP190102215 and DP210103706. WK is supported by the Pawsey Supercomputing Centre through the
Pawsey Centre for Extreme Scale Readiness (PaCER) program.

\bibliography{VortexConfinement}

\nolinenumbers

\end{document}